# Data Management for Context-Aware Computing

Wenwei Xue, Hungkeng Pung, Wenlong Ng
*School of Computing*
*National University of Singapore*
*3 Science Drive 2, Singapore 117543*
*{dcsxw, dcsphk, dcsnwl}@nus.edu.sg*

Tao Gu
*Institute for Infocomm Research*
*21 Heng Mui Keng Terrace, Singapore 119613*
*tgu@i2r.a-star.edu.sg*

**Abstract**

*We envisage future context-aware applications will dynamically adapt their behaviors to various context data from sources in wide-area networks, such as the Internet. Facing the changing context and the sheer number of context sources, a data management system that supports effective source organization and efficient data lookup becomes crucial to the easy development of context-aware applications. In this paper, we propose the design of a new context data management system that is equipped with query processing capabilities. We encapsulate the context sources into physical spaces belonging to different context spaces and organize them as peers in semantic overlay networks. Initial evaluation results of an experimental system prototype demonstrate the effectiveness of our design.*

## 1. Introduction

Context-aware computing is a key paradigm of pervasive computing in which applications monitor changes of context data in the environments and adapt operations to these changes in an unattended fashion [2]. As defined by Dey [3], *context* data is any information used to characterize the situation of an entity. We call such an entity a *context source*.

A context source can be an object or a space in the real world such as a person or a shop, as well as a logical component in the virtual world such as an application or a service [5]. A single source can provide many kinds of context data. For instance, a shop may provide its location, opening hours and product info.

A context-aware application often resides on some mobile device of a user [1][11][13][14][15], e.g., phone, PDA or laptop. To automatically recognize and react to the changing environment, the application must access context data from different sources. It should allow the user to specify which context sources to acquire data from rather than restricting the data access to a given set of sources determined by the application developer at built time. Furthermore, due to the dynamic nature of the context data, e.g., the location of a person or the crowd level in a shop, the application must be able to acquire data from a source in real-time. Otherwise the collected data may be obsolete and become useless.

As an example, consider a personalized shopping application running on the PDA of a user. When the user visits a shop and finds a product of interest, the application may acquire the prices of the product from nearby shops and recommend whether the user should buy the product from this shop. In order to make a good recommendation, the application requires several other context data of the shops including location, crowd level and reputation. What kind of shop to acquire data from and what kind of data to be acquired will vary depending on the shop that the user is visiting and the product the user wishes to buy. Context data from sources other than the shops may also influence the recommendation outcome, for example, the preferences of the user's family members and the schedules of the public transportations to the candidate shops. All data must be collected on-the-fly to make a timely reasoning.

Generally, the various types of context sources that a real-life context-aware application needs to acquire data from are likely to be distributed throughout wide-area networks like the Internet. Some of these sources are static, e.g., shops and homes, while the others could be mobile, e.g., persons and vehicles. It is therefore essential to organize these sources in an effective way and provide an efficient data lookup mechanism across them to ease application development.

Addressing this issue, in this paper we propose the design of a Context Data Management System (CDMS) for pervasive computing. The goal of the system is to facilitate the development of various context-aware applications as well as to optimize their acquisition and utilization of different data from multiple context sources. Figure 1 depicts the overall architecture of the CDMS.

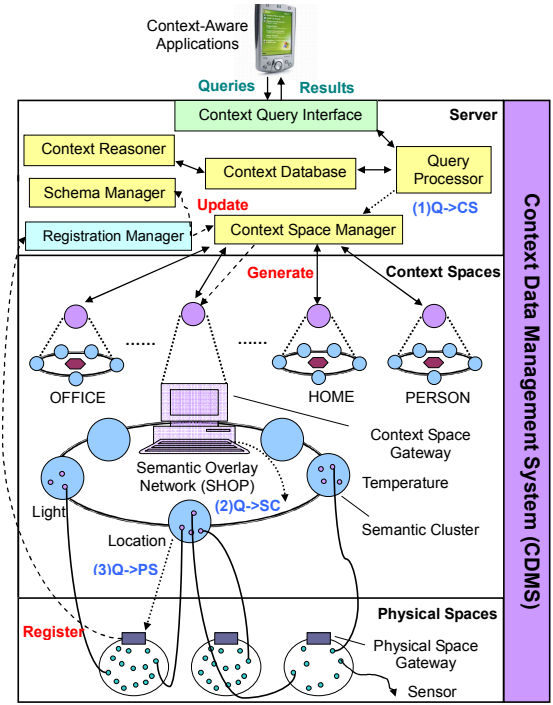

**Figure 1. Architecture of CDMS**

In our system, we represent and encapsulate context data using the concepts of context and physical spaces. We loosely define a *context space*, or a *context domain*, as a class of context sources that provides similar kinds of data. Examples of context spaces include PERSON, HOME and SHOP. We define a *context attribute* as a kind of context data provided by a context space, such as *name, location* of PERSON and *temperature, light* of SHOP. A *physical space* is a context source on which a gateway module specialized for data communication in our system has been installed.

To enable efficient context data lookup over multiple physical spaces, our CDMS first clusters these spaces into a set of context domains. The clustering process is automatic and the domains are continuously updated based on schema templates submitted by the physical spaces upon registration. The system then organizes all physical spaces sharing a common attribute in a context domain to form a semantic peer-to-peer (P2P) network, which we call a *semantic cluster*. The semantic clusters corresponding to different attributes in a domain are interconnected using a ring topology [6]. The P2P technology enables the autonomous joining and leaving of physical spaces in the system.

The CDMS provides a SQL-based query interface for context-aware applications to collect context data and subscribe events from various physical spaces. A hierarchical context reasoning scheme is employed in the system to deduce high-level context attributes from low-level ones. Historical context data can be stored at the system server or individual physical space gateways.

There has been much prior research work [1][2][5][8][9][11][13][15] on building context-aware systems to support application development. To the best of our knowledge, none of them has dealt with the organization and data lookup over physical spaces in multiple context domains as our proposed CDMS. We have implemented an experimental prototype of our system and measured its performance. Preliminary results demonstrate that the prototype achieved good query accuracy and response time with a simple Gnutella [4] P2P protocol.

The remainder of the paper is organized as follows. We describe the logical components of a physical space gateway in Section 2. We present the organization of physical spaces and the processing of context queries in our system in Section 3. Section 4 presents preliminary performance evaluation results of our system prototype. Section 5 discusses the related work. We conclude the paper with future research directions in Section 6.

## 2. Gateways of Physical Spaces

Our CDMS consists of a system server and a number of physical space gateways. The main functional units of the system are located at the server, as shown in the upper part of Figure 1. The server can be implemented using a single powerful machine, or be decentralized as a sever cluster.

Each physical space in our system has a *physical space gateway* (*PSG*) that acts as the communication bridge between the space and the server. The gateway is a logical module running at any computer of choice in the physical space. For instance, the gateway of a person space can be located at the person's PDA, and a home space gateway can be the owner's PC in the house.

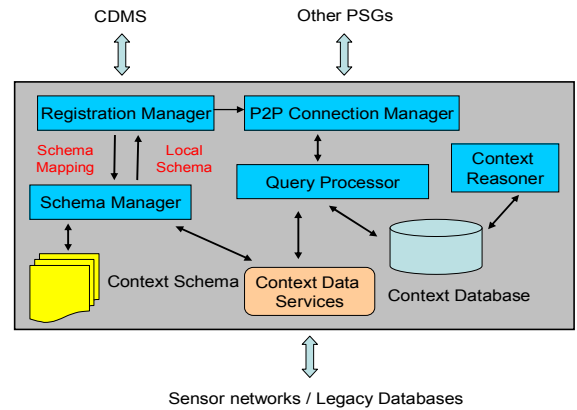

**Figure 2. Components of a PSG**

The gateway provides a single and uniform interface for the system server and the other gateways to acquire context data from the sensor network or legacy database contained in a physical space. It also provides a set of functional components to manage and manipulate the context data in the space. The component diagram of a PSG is illustrated in Figure 2. We describe individual components in order as follows.

- **Context Schema**. In our prototype, we use a simple attribute-value model [2] for context data representation at both a PSG and the server. The purpose is to make the initial design of system components generic while extensible to support more expressive context models, e.g., ontologies [5], in the future.

Given this model, the *context schema* of a PSG is a description of the context domain the physical space belongs to and the context attributes the space provides in this domain. It includes various metadata such as the names of the context domain and attributes, the attribute data properties, whether the domain is a subset of the other domain, and so on. The schema manager of a PSG maintains its context schema.

- **Registration Manager**. A PSG must register with system server before its data can be searched and used by the applications running on top of the CDMS. The registration manager of a PSG performs the task. It sends a registration request to its counterpart at the server, as shown in Figure 1. The request contains the IP address and schema template of the PSG. The template is an XML representation of the context schema.

The context schema of a PSG may change from time to time due to the addition/removal of sensors or data objects in the physical space. In this case, the registration manager will send an update request to the server and inform it about such changes.

- **P2P Connection Manager**. Each PSG registered to the CDMS joins as a peer in a number of semantic P2P networks maintained by the system server. The P2P communication protocols are implemented in the P2P connection manager.

- **Context Data Services**. We specify a set of context data services at a PSG that encapsulates various kinds of context data acquisition, such as collection, filtering, aggregation and reasoning. These services are the only methods for the PSG to acquire context data from the sensor network or legacy database in the physical space.

Our specification of context data services defines the input/output behaviors rather than the implementation details of the services. Before a PSG registers to the CDMS, all these services must be implemented and deployed at the PSG. The implementation of a service at a PSG is dependent on the specific sensor hardware and database APIs used in the physical space.

We have designed the services to incorporate many typical features of data acquisition required in context-aware computing. These include pull or push-based, one-time or continuous, blocked or unblocked context data acquisition as well as quality and privacy control of context data. We also allow applications to subscribe to special *events* in a physical space, such as *isFire* and *isEating*, which are deduced attributes with "true" and "false" semantics provided by a local context reasoner.

- **Query Processor**. When a PSG receives a context query routed from a P2P neighbour, the query is parsed and executed by a local query processor. If the query condition is satisfied at this moment, the PSG acquires the query result and returns it directly to the system server. All context attribute values required for query execution are obtained by invoking the corresponding context data services at the PSG. If the query requests the storage of certain period of data, the query processor puts such data into a context database at the PSG.

We have developed a PSG toolkit in the prototype implementation of our CDMS. After downloading and installing the toolkit, any PC in a wide-area network can configure itself as a PSG using the toolkit GUI and register as a member in our system.

## 3. Context Query Processing

The main functionality of the CDMS is to provide effective organization of physical spaces and efficient processing of context queries for data acquisition over these spaces. In this section, we present a number of techniques we design in our system to support context query processing.

### 3.1. Query Interface

We have designed and implemented a SQL-based query interface in our system prototype. Context-aware applications or services can issue declarative context queries to acquire context data from physical spaces via this interface. Previous work [7] has suggested that SQL-based context query language is as effective as complex counterparts like RDQL in many aspects.

Two classes of queries are supported in our context query language. They are called *data collection queries* and *event subscription queries*. Queries 1 and 2 give examples of the two query classes.

*Query 1*:

```
SELECT  friend_list  FROM  PERSON
WHERE   name = "Keith"
```

*Query 2*:

```
SUBSCRIBE  isVacant  FROM  OFFICE
WHERE      location = "S14 #06-20, NUS"
```

A data collection query returns the values of a list of context attributes at physical spaces that satisfy the query condition. An event subscription query subscribes the application that issues the query to events in a physical space. A CONT keyword can be used to specify that a data acquisition query is continuous and push-based, as exemplified by Query 3. In this case, the sample period and duration of the continuous acquisition are specified using the SAMPLE PERIOD and LIFETIME clauses.

*Query 3*:

SELECT   CONT   location   FROM   PERSON
WHERE    name = "Keith"
SAMPLE PERIOD   1 min   LIFETIME   2 hours

### 3.2. Schema Matching

Applications issue context queries to the declarative interface based on a global set of context schemas that describe various context domains. These schemas are maintained by the schema manager at the system server (Figure 1). For instance, "PERSON" in Query 1 refers to the global schema of the PERSON domain.

Our CDMS aims to support autonomous physical spaces owned by multiple organizations. It is impractical to impose a common name space of context attributes or domains among all spaces for schema composition. Instead, we allow each space to compose its context schema independently while we publish current global schemas via a public web interface for reference only.

With this system setting, an immediate problem we face is to integrate the local schemas submitted from physical spaces into the global schemas at the server. Note that two spaces can specify different names, e.g., *name* and *personName*, for an attribute with the same semantics; they can also specify different names, e.g., *HOME* and *HOUSE*, for a domain with the same set of attributes.

To solve this problem of context schema matching [16], we have designed a name-based schema matcher as a core subcomponent of the schema manager. The matcher uses a Context Attribute Matching algorithm to match the attributes in a local schema with those in the global schemas. The algorithm applies a number of linguistic matching criteria sorted in a decreasing order of weights, such as stemming, substring and synonym detection. The weights of the criteria are dynamically adjusted based on the recent matching accuracy they achieve, which makes the algorithm adaptive to current patterns of schema inputs. In addition, the matcher uses a Context Schema Matching algorithm to try to integrate a local schema into one of the global schemas based on the largest common subset of matched attributes.

For accuracy improvement, the matcher reports all candidate matches of attributes/schemas it finds to the system administrator for confirmation via a GUI. New attributes/schemas that could not be matched with any existing ones are added into the global schemas. The mapping between a local schema and the global schemas is sent to and stored at the PSG (see left of Figure 2). When a query specified upon the global schemas is routed to a PSG, the query syntax is first converted to the local schema before the query is evaluated.

As a whole, the schema matcher helps to integrate all local schemas from physical spaces into a set of global schemas at the server, based on which context queries to our system are syntactically composed and issued. The initial set of global schemas can be pre-defined or empty. More details about the matcher design are available from our previous paper [16].

### 3.3. Semantic P2P Overlay

Each context domain, after it is first generated in the schema manager, is mapped to a logical one-dimensional ring topology that contains several semantic clusters, as shown in Figure 1. The ring forms a semantic overlay network that is an extension to prior work of Gu et al. [6]. Each semantic cluster in the ring corresponds to a context attribute in the domain.

A semantic cluster is implemented as a P2P network in which each peer is a PSG. As a physical space may not provide all kinds of attributes in the current global schema of the context domain, its PSG will only join those clusters for attributes it has. As an initial solution, we use Gnutella [4] version 0.4 as the P2P protocol in our system prototype.

The rings for all context domains are created and maintained by the context space manager at the server. A *Context Space Gateway* (*CSG*) is created as a special cluster in each ring as the entrance for query routing. The CSG connects into the ring as any other cluster while it is actually a subcomponent of the context space manager rather than a P2P network. It maintains the ring topology and is responsible for the generation of new semantic clusters when the context domain evolves.

A PSG can leave the semantic clusters it joins at any time. The leaving is automatically detected by the P2P protocol at the neighboring peers. The PSG must send a new registration request in order to rejoin the system.

### 3.4. Query Evaluation Plan

When the query interface receives a context query, the query string is forwarded to a query processor at the server. The processor parses the query and examines whether its syntax is correct given the global schemas. A query with wrong syntax is not further processed and an error message is returned to the application.

The query processor generates an evaluation plan [10] for a context query that is successfully parsed. The plan is a tree of operators in which data is streamed from bottom to up. A query plan in our CDMS has two main differences with a traditional plan in a relational DBMS. First, in the plan a scan operator is created for each context domain specified in the FROM clause of the query. The operator encapsulates the lookup process of context data in the domain rather than that from a disk table. Second, the query condition is evaluated at individual PSGs. This means the selection operators are lower than the scan operators in the query plan.

The process of context data lookup in a scan operator includes three separate steps of mapping the context query to different data structures in the CDMS. This is illustrated by the flow (1)-(3) in Figure 1.

First, the scan operator searches an index of context domains in the context space manager to find the CSG of the domain. Next, the operator establishes connection and sends a lookup request to the CSG. The request is a parsed version of query string that contains the query conditions and the attribute list to acquire. The request is injected into the ring by the CSG and is processed by the first semantic cluster corresponding to one attribute in the projection list. Finally, the request is disseminated in the P2P network. Each PSG that satisfies the query condition reports data directly to the server-side scan operator. To cope with the Gnutella flooding protocol in our prototype, a *TTL* (*Time-To-Live*) value is set to each query. It indicates the number of hops the lookup request of a query can reach in the P2P network.

### 3.5. Hierarchical Context Reasoning

The CDMS is equipped with a hierarchical and hybrid reasoning scheme that deduces high-level context events from low-level attributes or events. The scheme consists of three-level context reasoners: (i) intra-space reasoners at the PSGs, (ii) intra-context domain reasoners at the CSGs, (iii) an inter-domain reasoner at the server. We apply machine learning techniques for the lower two levels while rule-based techniques for the highest level. The output of the context reasoners can be stored at the context databases at various locations and accessed by queries. We omit the detailed reasoning techniques here.

## 4. Performance Evaluation

We have implemented a prototype of our CDMS in Java and deployed it in our research lab. We present preliminary performance evaluation results measured from the prototype in this section.

### 4.1. Experimental Setup

We used one desktop PC for the system server and four PCs to virtualize the PSGs in the experiments. Each PC has an Intel Pentium IV 2GHz CPU and 1GB main memory running Windows XP. The physical spaces involved in the experimental prototype were randomly picked from one of three context domains. Each space had 30 context attributes that were randomly picked from a common set of attributes in the domain.

We registered 200-1000 PSGs to one *query cluster* in the semantic P2P networks of the three domains. A context query was sent to this cluster and executed on the PSGs in the cluster in each run of our experiments. The position of the query cluster was randomly picked in a run. The numbers of PSGs in other clusters were random. Each value in the following figures was the average of tens of independent experimental runs.

All PSGs in the query cluster were virtualized by the PSG processes in the four PCs uniformly. We have tried to use more PCs for the query cluster of the same scale and found that the performance difference was little.

### 4.2. Timing of Physical Space Registration

Figure 3 shows the time breakdown in the prototype upon the registration of a physical space. In the figure, "registration request" is the time for the PSG to send a request to the system sever and the time for the server to parse the request; "schema matching" is the time spent in the schema manager to incorporate the local PSG schema to the global schemas; "return SC list" is the time for the context space manager to identify all semantic clusters the PSG needs to join and return this list to the PSG, including the time to create new semantic overlays or clusters if necessary; "P2P connection establishment" is the time the PSG takes to connect to its P2P neighbors.

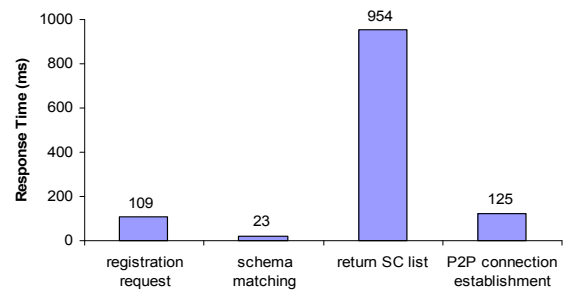

**Figure 3. Time breakdown of physical space registration**

In the figure we see that, the total PSG registration time was about 1.2 second in our system prototype. It suggests that a PSG can join the system rapidly. The time of "return SC list" was relatively large because the computation overhead to lookup and create semantic data structures for the PSG to join was considerable. The time for the other items was small in comparison.

### 4.3. Effect of Query Processing

In this section, we tested the effectiveness of context query processing using two performance metrics: (i) the *response time* of a query that is the time taken in all system components to process the query, (ii) the *recall* (*accuracy*) of a query that is the percentage of PSGs with the required query results that are successfully identified by the semantic P2P search.

Figure 4 shows the time breakdown of processing a context query with selection and projection operations. In this experiment, there were 1000 PSGs in the query cluster a random percent of which had required query results in different experimental runs. The TTL value of the query was set to be 8.

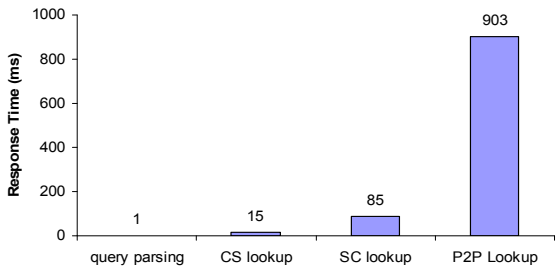

**Figure 4. Time breakdown of context query processing**

As expected, the figure shows that the time for P2P query dissemination and evaluation dominated the entire query response time, although the value of 0.9 seconds is still acceptable in practice. The time spent on query parsing, context space and semantic cluster lookup was tiny and negligible. The large P2P search time was due to the implementation of a simple Gnutella protocol in the prototype.

We next varied the TTL value of a context query in our prototype to study how this parameter affects the query accuracy. A small TTL value suggests a shorter query response time for the applications, but the query may not be able to reach enough qualifying PSGs in the P2P networks and its accuracy will suffer. The results are shown in Figure 5. We used 1000 PSGs for the query cluster in this experiment.

The figure illustrates that as long as the TTL value was not set to be very small (< 4 in this experiment) the query accuracy was reasonably good. Moreover, a TTL value that was moderate (6 in the experiment) could help to converge query accuracy to about 100%.

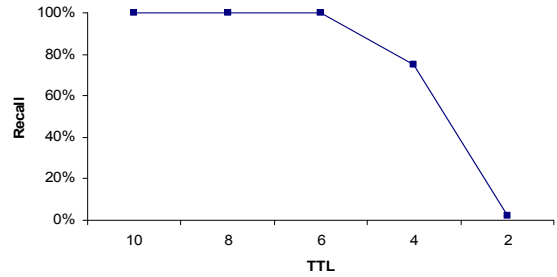

**Figure 5. Query accuracy with different TTL values**

Finally, we varied the number of PSGs in the query cluster to examine how long a context query can finish its processing in the prototype after all PSGs having required query results were found. The TTL value of a query was set to be 8 in this experiment. This value guaranteed 100% query accuracy as shown in Figure 5. The results are given in Figure 6. It indicates that the query response time increased with size of P2P network whereas the increase curve was smooth and non-linear.

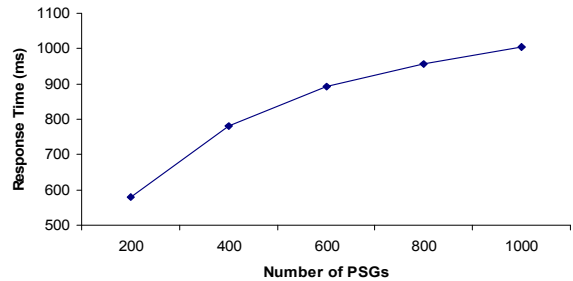

**Figure 6. Query response time with different network sizes**

### 5. Related Work

There is little previous work in context-aware computing that considers multiple physical spaces in different context domains. Most existing systems only involve a single physical space, or multiple spaces in a single domain. This state of the art proves the design of our CDMS is novel and promising.

CoBrA [1] applies Semantic Web technologies for context acquisition and reasoning in a physical space. CIS [8] defines physical spaces as one class of entities that provide context data; the other entities are people, devices and networks. Our definition of physical spaces includes all types of context sources registered to the CDMS. The "places" in CoolTown [9] are similar to physical spaces whereas all places in the system are managed as a whole without considering the high-level semantics of individual context domains. In Contory [11], a single context space that contains the set of all available attributes is defined. None of these systems has investigated concurrent context data management over spaces in multiple context domains as our system.

A physical space in our study is similar to an active or smart space in Gaia [13]. Gaia provides services to manage and program a physical space and abstracts data lookup into a context service. Data can be acquired from a context provider in Gaia by issuing a query or by subscribing to the event channel published by the provider. We organize the physical spaces as peers in several semantic overlay networks and map context data lookup to P2P lookup in these networks.

Roman et al. [12] proposed a declarative approach based on view definitions over tuple spaces to cooperate operations of application agents on host computers. The tuple spaces and views in this work bear a similarity to the context spaces and physical spaces in our work, respectively. Shilit [14] proposed an infrastructure for context-aware systems consisting of distributed context servers. The servers maintain information about context domains while context data from physical spaces of all domains are stored at a central server. In comparison, context data in our system are distributedly located in various physical spaces.

SOCAM [5] and Semantic Space [15] have studied the context data management over physical spaces in different domains. However, the semantics of "context space" is not utilized to organize the physical spaces in these systems as in our CDMS. Each physical space separately registers its own context wrapper for every context provider in the space.

## 6. Conclusion

We propose a data management system for context-aware computing and present our system design in this paper. We develop a gateway module as the uniform interface to acquire context data from lots of physical spaces in various context domains. We use semantic P2P networks to organize physical spaces having the same kinds of context attributes, and propose techniques for processing context queries for data acquisition. Initial evaluation results of the system prototype show that our design achieves good query response time and accuracy.

Our future work includes evaluating the performance of our system with more efficient P2P protocols, and designing P2P protocols that accelerate the lookup of dynamic context data.

## 7. Acknowledgements

This work is funded by the Science and Engineering Research Council of Singapore under the research grant no. SERC 0521210083.

## 8. References

[1] H. Chen, T. Finin, and A. Joshi, "Semantic Web in a Pervasive Context-Aware Architecture", *Proc. AIMS*, 2003, pp. 33-40.

[2] G. Chen, and D. Kotz, "A Survey of Context-Aware Mobile Computing Research", Technical Report TR20 00-381, Department of Computer Science, Dartmouth College, 2000.

[3] A.K. Dey, "Understanding and Using Context", *Personal Ubiquitous Computing*, vol. 5, no. 1, 2001, pp. 4-7.

[4] Gnutella, http://rfc-gnutella.sourceforge.net/.

[5] T. Gu, H.K. Pung, and D. Zhang, "A Service-Oriented Middleware for Building Context-Aware Services", *Journal of Network Computer and Applications*, vol. 28, no. 1, 2005, pp. 1-18.

[6] T. Gu, H.K. Pung, and D. Zhang, "A Semantic P2P Framework for Building Context-aware Applications in Multiple Smart Spaces", *Proc. EUC*, 2007, pp. 553-564.

[7] P.D. Haghighi, A. Zaslavsky, and S. Krishnaswamy, "An Evaluation of Query Languages for Context-Aware Computing", *Proc. DEXA*, 2006, pp. 455-462.

[8] G. Judd, and P. Steenkiste, "Providing Contextual Information to Pervasive Computing Applications", *Proc. PerCom*, 2003, pp. 133-142.

[9] T. Kindberg, and J. Barton, "A Web-Based Nomadic Computing System", *Computer Networks*, vol. 35, no. 4, 2001, pp. 443-456.

[10] R. Ramakrishnan, and J. Gehrke, *Database Management Systems*, McGraw-Hill, Columbus, OH, USA, 2002.

[11] O. Riva, and C. Flora, "Contory: A Smart Phone Middleware Supporting Multiple Context Provisioning Strategies", *Proc. SIUMI*, 2006, pp. 68.

[12] G.C. Roman, C. Julien, and A.L. Murphy, "A Declarative Approach to Agent-Centered Context-Aware Computing in Ad Hoc Wireless Environments", *Proc. SELMAS*, 2005, pp. 94-109.

[13] M. Román, C. Hess, R. Cerqueira, A. Ranganat, R.H. Campbell, and K. Nahrstedt, "A Middleware Infrastructure for Active Spaces", *IEEE Pervasive Computing*, vol. 1, no. 4, 2002, pp. 74–82.

[14] B.N. Shilit, "A Context-Aware System Architecture for Mobile Distributed Computing", PhD Thesis, Department of Computer Science, Columbia University, 1995.

[15] X. Wang, J.S. Dong, C.Y. Chin, S. Hettiarachchi, and D. Zhang, "Semantic Space: An Infrastructure for Smart Spaces", *IEEE Pervasive Computing*, vol. 3, no. 3, 2004, pp. 32-39.

[16] W. Xue, H.K. Pung, P.P. Palmes, and T. Gu, "Schema Matching for Context-Aware Computing", *To Appear in Proc. Ubicomp*, 2008.